\documentclass[11pt,onecolumn]{article} 
\usepackage{latex8}
\usepackage{times}

\usepackage{algorithm}
\usepackage{algpseudocode}
\usepackage{graphicx}
\usepackage{amsmath}
\usepackage{amssymb}

\begin{document}

\title{Inexact Graph Matching Using Centrality Measures}


\author{
 Shri Prakash Dwivedi \thanks{\text{Email: shriprakashdwivedi@gbpuat-tech.ac.in}}\\
% \textit{Department of Information Technology}\\
% \textit{G. B. Pant University of Agri. \& Tech.}\\
}

\maketitle

\begin{abstract}
Graph matching is the process of computing the similarity between two graphs. Depending on the requirement, it can be exact or inexact. Exact graph matching requires a strict correspondence between nodes of two graphs, whereas inexact matching allows some flexibility or tolerance during the graph matching. In this chapter, we describe an approximate inexact graph matching by reducing the size of the graphs using different centrality measures. Experimental evaluation shows that it can reduce running time for inexact graph matching.%\footnote{Email: shriprakashdwivedi@gbpuat-tech.ac.in}
%\keywords{Graph matching, graph edit distance, centrality measures, structural pattern recognition}
\end{abstract}

\section{Introduction}
Graph Matching (GM) is one of the important research areas in graph-based representation in structural pattern recognition. GM is the process of computing the similarity between the two graphs. Depending on the nature of the matching, it has been broadly classiﬁed into two varieties, exact GM and inexact or error-tolerant GM. For exact GM, strict correspondence is necessary between each node and edge of the ﬁrst graph to the corresponding nodes and edges of the second graph. Exact GM is like a graph isomorphism problem in which a bijective mapping is required from the nodes of the ﬁrst graph to the nodes of the second graph.

Exact GM although theoretically appealing, may not be useful in many real-world applications, as due to the existence of noise or distortion during the processing, the input graph data may be altered. In such situations, we use inexact GM or error-tolerant GM due to its flexibility to accommodate errors during the process of matching. Polynomial time solution for GM is not available. Graph isomorphism problem is neither shown to be in $NP$-complete nor in $P$. On the other hand, the subgraph isomorphism problem is known to be $NP$-complete. Due to non-availability of exact polynomial time algorithms for GM problem, several approximation algorithms and heuristic have been proposed.

An extensive survey of different GM methods is explained in \cite{Conte 2004} and \cite{Foggia et al. 2014}. In \cite{Bunke 1998} author describes a precise framework for inexact GM. A-star search technique for finding minimum cost paths is described in \cite{Hart et al. 1968}. Inexact GM of the attributed relational graphs (ARG) is described in \cite{Tsai and Fu 1979}. In \cite{Sanfeliu and Fu 1983} authors introduced a distance measure for non-hierarchical ARG by considering the cost of recognition of nodes.

A category of GM algorithms utilizing the spectral technique of algebraic graph theory has been introduced \cite{Caelli and Kosinov 2004}, which depend on the fact that adjacency matrices of similar graphs will have a similar decomposition \cite{Robles-Kelly and Hancock 2005}, \cite{Shokoufandeh et al. 2005}.

%A survey of theoretical results and a formal framework for error-tolerant GM is described in \cite{Bunke1998}. Heuristic information from the problem domain is used to provide a formal mathematical theory for graph searching in \cite{Hartetal1968}. The paper \cite{TsaiFu1979} describes an ordered-search algorithm for determining error-correcting isomorphism. A technique to determine a distance measure between two non-hierarchical attributed relational graphs is described in \cite{SanfeliuFu1983}.

%In the paper \cite{CaelliKosinov2004}, the authors describe how the graph eigenspaces can be used for solving error-tolerant GM. Graph spectral seriation method is used to convert adjacency matrix into a string; then the GM problem is posed as a maximum a posteriori probability alignment of the seriation sequences for the pair of graphs \cite{KellyHancock2005}. A framework for indexing hierarchical image structure utilizing graph spectra, which embeds the topological structure of a directed acyclic graph \cite{Shokoufandehetal2005}.

A novel class of GM techniques based on the so-called graph kernel, which uses the concept of kernel machines to graph domain, is described in \cite{Gartner 2008}, \cite{Neuhaus and Bunke 2007}. Graph kernel enables us to utilize statistical pattern recognition methods to the structural pattern. The essential types of graph kernels are convolution kernel, diffusion kernel and random walk kernel \cite{Haussler 1999}, \cite{Lafferty and Lebanon 2005}.

Another technique of GM is based on geometric graphs in which each vertex has its associated coordinate in two-dimensional space \cite{Dwivedi and Singh 2018a}, \cite{DwivediSingh2019}. Geometric GM using the edit distance approach is demonstrated to be $NP$-hard in \cite{Cheong et al. 2009}. Geometric GM using a probabilistic approach is described in \cite{Armiti and Gertz 2014} and in the paper, \cite{Pinheiro et al. 2017} authors have presented geometric GM based on Monte Carlo tree search.

Graph Edit Distance (GED) is one of the important techniques used for inexact GM \cite{Bunke and Allerman 1983}, \cite{Sanfeliu and Fu 1983}. GED between two graphs is defined as the minimum edit operations needed to convert the first graph into another one. GED is the generalization of string edit distance. Exact algorithms for GED are computationally expensive and are exponential on input graphs' size. To make GED computation feasible, many approximate methods using local search, greedy method, neighborhood search, bipartite GED, homeomorphic, GED etc. have been proposed \cite{Dwivedi and Singh 2017},\cite{Ferrer et al. 2015},\cite{Neuhaus et al. 2006},\cite{Riesen and Bunke 2009},\cite{Riesen and Bunke 2015},\cite{Sorlin and Solnon 2005}, \cite{DwivediSingh2020}, \cite{Dwivedi2019}.

In \cite{Dwivedi and Singh 2018}, the authors proposed an approach to inexact GM by contracting the nodes from the graphs based on their degree centrality. In this chapter, we describe this approach to perform inexact GM by reducing the size of the graphs using different centrality measures such as eigenvector, betweenness and PageRank centrality. It leads to a reduction in search space needed to compute GED between two graphs. We perform the experimental evaluation to demonstrate that these centrality measures can be used as a trade-off for running time and accuracy for GM algorithms.

This chapter is organized as follows. Section 2, contains preliminaries and motivation. Section 3, presents inexact GM using centrality measures. Section 4, describes the experimental evaluation and finally section 5, contains the conclusion.

\section{Preliminaries and motivation}
In this section we explain the basic definitions related to GM. For a detailed description the reader is referred to \cite{Aggarwal and Wang 2010},\cite{Neuhaus and Bunke 2007}. We also describe the motivation for our work.

A \textit{graph} $G$ is defined as $G= (V,E,\mu,\nu)$, where $V$ is the set of vertices, $E$ is the set of edges, $\mu$ is a node labeling function $\mu: V \rightarrow L_V$, and $\nu$ is edge labeling function $\nu: E \rightarrow L_E$. Here, $L_V$ is the node label set and $L_E$  is the edge label set.
A graph $G_1$ is called \textit{subgraph} of another graph $G_2$, when $V_1 \subseteq V_2$;  $E_1 \subseteq E_2$; for each node $u$ of graph $G_1$, we have $\mu_1(u)=\mu_2(u)$; similarly, for each edge $e$ of $G_1$, we have $\nu_1(e)=\nu_2(e)$.

A sequence of edit operations that convert one graph $G_1$ to another graph $G_2$ is called as \textit{edit path} from $G_1$ to $G_2$. The simple edit operations include insertion, deletion and substitution nodes and edges. Insertion and deletion of node $u$ is denoted respectively by $\epsilon \rightarrow u$ and $u \rightarrow \epsilon$, whereas substitution of node $u$ by node $v$ is denoted by $u \rightarrow v$. Similarly insertion and deletion of edge $e$ is represented respectively by $\epsilon \rightarrow e$ and $e \rightarrow \epsilon$, while substitution of edge $e$ by edge $f$ is denoted by $e \rightarrow f$.

The \textit{GED} between two graphs $G_i = (V_i, E_i, \mu_i,\nu_i)$ for $i=1,2$ is defined by $$ GED(G_1, G_2)= min_{(e_1,...,e_k) \in \varphi (G_1, G_2)} \sum_{i=1}^{k} c(e_i)$$ where $c(e_i)$ is the cost of edit operation $e_i$ and $\varphi (G_1, G_2)$ represents the set of all edit path transforming $G_1$ to $G_2$.

\textit{Node contraction} is the process of deleting nodes and its associated edges provided it is not a cut vertex \cite{Dwivedi and Singh 2018}. \textit{$k$-degree node contraction} on a graph $G$ is the process of contracting all nodes of degree $k$ in graph $G$.
\textit{$k^{*}$-degree node contraction} is the task of applying $k$-degree node contraction iteratively on a graph $G$, from $i=1$ to $k$. \textit{$k^{*}$-GED} is defined as GED between $G_1$ and $G_2$, with $k^{*}$-degree node contraction applied on both $G_1$ and $G_2$.

The applications of exact GM to real-world applications is rather limited due to the presence of noise or error during the processing of the graphs. Inexact GM offers an alternative to perform approximate GM. Due to exponential complexity associated with GED, other methods have been introduced to perform efficient GM at the cost of a slight decrease in accuracy. A technique proposed in \cite{Dwivedi and Singh 2018} is based on removing the nodes based on their degree centrality to decrease the size of the matching graphs. However, degree centrality may not always be the best criteria to ignore the nodes. Depending on the structure and properties of the different dataset, we can select the appropriate centrality measure to delete the nodes for reducing the size of the graphs. In this chapter, we use eigenvector, betweenness and PageRank centrality in addition to degree centrality to reduce the size of the graphs for estimating an early approximate GM between two graphs.

Now we briefly explain the above centrality measures \cite{Newman 2010}. Centrality of the node in a graph signifies its relative importance in the graph. The centrality measures aim to find the most important or central nodes of a graph or network. Simplest centrality measure is \textit{degree centrality}, which simply refers to the degree of the given node. A node with more adjacent nodes or neighbors will have higher degree centrality as compared to nodes with a fewer connection. \textit{Betweenness} centrality of a node is based on the extent by which this node lies on the paths between other nodes. \textit{Eigenvector} centrality is a generalization of degree centrality, which assigns each node a value proportionate to the sum of the values of its neighbors. For a node $u_i$ its eigenvector centrality is given by $x_i=\kappa_1^{-1}\sum_j A_{ij} x_j$, where $\kappa_1$ is the largest eigenvalue of adjacency matrix $A$ and $A_{ij}$ is an element of $A$. In PageRank centrality, the centrality of a node is proportionate to the centrality of its neighbors divided by their outgoing degree. The \textit{PageRank} centrality is defined by $x_i = \alpha \sum_j A_{ij} \frac{x_j}{k_j} + \gamma$, where $\alpha$ is a free parameter, $k_j$ is the outgoing degree and $\gamma$ is a constant.

\section{Inexact graph matching}
To reduce the computation time of inexact GM, we ignore the nodes from the graphs with less centrality value before computing a similarity score using GED between two graphs. \\

\textbf{Definition 1.}
 $t$-centrality node contraction is the process of contracting $t$ nodes from a graph $G$ with least centrality values of a given centrality measure.\\

The above definition implies that starting from the node with the lowest centrality value in a graph $G$, up to $t$ nodes are deleted provided they are not a cut vertex. Depending on the centrality measure used $t$-centrality node contraction ($t$-NC) can be $t$-degree centrality node contraction ($t$DC-NC), $t$-betweenness centrality node contraction ($t$-BC-NC), $t$-eigenvector node contraction ($t$-EV-NC) and $t$-PageRank node contraction ($t$-PR-NC). \\

\textbf{Definition 2.}
 $t$-degree centrality node contraction is the operation of contracting $t$ nodes of the smallest degree from a graph $G$. \\

When $t$ is equal to the number of nodes of degree $k$ in a graph, then $t$-degree centrality node contraction corresponds to $k$-degree node contraction. \\

\textbf{Definition 3.}
 $t$-betweenness centrality node contraction is the operation of contracting $t$ nodes with the lowest betweenness score from a graph $G$.\\

\textbf{Definition 4.}
 $t$-eigenvector centrality node contraction is the process of contracting $t$ nodes with the lowest eigenvector centrality from a graph $G$.\\

\textbf{Definition 5.}
 $t$-PageRank centrality node contraction is the process of contracting $t$ nodes with the lowest PageRank score from a graph $G$. \\

\textbf{Definition 6.}
 $t$-centrality GED computation between two graphs $G_1$ and $G_2$ is defined as GED between these graphs, when $t$ nodes of least centrality of both graphs $G_1$ and $G_2$ have been contracted. \\

In the above definition depending on the actual centrality criteria used $t$-centrality GED computation ($t$-GED) corresponds to $t$-degree centrality GED computation ($t$-DC-GED), $t$-betweenness centrality GED computation ($t$-BC-GED), $t$-eigenvector GED computation ($t$-EV-GED) and $t$- PageRank GED computation ($t$-PR-GED).

\subsection{Edit Cost}
We can define the edit cost of $t$-GED by using an additional operation $c(u \rightarrow \epsilon)= 0$, for $t$ vertices of the graph $G$ having the lowest score of the given centrality measure.

$t$-GED utilizes the Euclidean distance and allocates the constant cost to insertion, deletion and substitution of vertices and links. For two graphs $G_1$ and $G_2$, having vertices $u \in V_1$, $v \in V_2$ and links $e \in E_1$, $f \in E_2$, we specify the extended edit cost function as given below. \\
$c(u \rightarrow \epsilon)= x_{node}$ \\
$c(\epsilon \rightarrow v)= x_{node}$ \\
$c(u \rightarrow v)= y_{node}.|| \mu_1(u) - \mu_2(v)||$ \\
$c(e \rightarrow \epsilon)= x_{edge}$ \\
$c(\epsilon \rightarrow f)= x_{edge}$ \\
$c(e \rightarrow f)= y_{edge}. || \nu_1(e) - \nu_2(f)||$ \\
$c(u \rightarrow \epsilon)= 0$, if $u$ is one of the $t$ nodes of the lowest centrality value and is not a cut vertex. \\
Here $x_{node}$, $y_{node}$, $x_{edge}$, $y_{edge}$ are positive constants.

\subsection{Algorithm}
The computation of inexact GM using $t$-centrality node contraction is outlined in Algorithm 1. The input to the $t$-Centrality-Graph-Edit-Distance algorithm is two graphs $G_1=(V_1, E_1, \mu_1,\nu_1)$, $G_2=(V_2, E_2, \mu_2,\nu_2)$ and a parameter $t$. The output to the algorithm is the minimum cost $t$-GED between $G_1$ and $G_2$. The algorithm calls the procedure $t$-Centrality-Node-Contraction in lines 1--2 for graphs $G_1$ and $G_2$ respectively to remove $t$ nodes having the lowest centrality value provided they are not cut vertex. $G_1'$ and $G_2'$ are the resultant graphs obtained after performing $t$-Centrality-Node-Contraction on $G_1$ and $G_2$ respectively, such that $V_1' = \{u_1',...,u_{n'}'\}$ and $V_2' = \{v_1',...,v_{m'}'\}$. Line 3 initializes an empty set $A$.	 The vertex $u_1'$ of $G_1'$ is substituted by each vertex $v_j'$ of $G_2'$ in the \textit{for} loop of lines 3--6, and deletion of $u_1'$ is performed in line 7. The computation of the minimum cost edit path is performed in the \textit{while} loop of lines 8--27. \textit{If} loop in line 10 check, whether $C_{min}$ is a complete edit path, so that it completely transform $G_1'$ to $G_2'$.  If all nodes $V_1'$ are processed (line 13), then remaining nodes of $V_2'$ are simply inserted in $C_{min}$ in \textit{for} loop of lines 14--16. Similarly, all unprocessed vertices of $V_1'$ is substituted by all vertices of $V_2'$ along with the deletion of vertices of $V_1'$ in the \textit{for} loop of lines 19--23, and $A$ is updated in line 24. \\

\begin{algorithm}
\caption{\bf :  $t$-Centrality-Graph-Edit-Distance $(G_1,G_2)$}
\begin{algorithmic}[1]
\Require  Two Graphs $G_1$, $G_2$, %where $G_i = (V_i, E_i, \mu_i,\nu_i)$ for $i=1,2$
                       where $V_1 = \{u_1,...,u_n\}$ and $V_2 = \{v_1,...,v_m\}$ and a parameter $t$
\Ensure  A minimum cost $t$-GED between $G_1$ and $G_2$

   \State $G_1' \leftarrow t$-Centrality-Node-Contraction $(G_1, t)$
   \State $G_2' \leftarrow t$-Centrality-Node-Contraction $(G_2, t)$
   \State $A \leftarrow \emptyset$
   \For {each $(v_j' \in V_2')$}
%   {
   \State $A \leftarrow A \cup \{ u_1' \rightarrow v_j' \}$
%   }
   \EndFor
   \State $A \leftarrow A \cup \{ u_1' \rightarrow \epsilon \}$
   \While { (True) }
%   {
%   \State Prune $A$ using optimizing techniques
   \State Compute minimum cost edit path $C_{min}$ from $A$
   \If {($C_{min}$ is a complete edit path)}
    \State \textbf{return} $C_{min}$
   %\Return $C_{min}$
   \Else
       \If {(all vertices $(u_i' \in V_1')$ are visited)}
        \For {all unvisited $(v_j' \in V_2')$}
%         {
         \State $C_{min} \leftarrow C_{min} \cup \{ \epsilon \rightarrow v_j' \} $
%          }
        \EndFor
        \State $A \leftarrow A \cup \{ C_{min} \}$
        \Else
       \For {(all unvisited vertices $(u_i' \in V_1')$)}
%       {
         \For {(each $(v_j' \in V_2')$)}
%         {
         \State $C_{min} \leftarrow C_{min} \cup \{ u_i' \rightarrow v_j' \} \cup \{ u_i' \rightarrow \epsilon \}$
%          }
         \EndFor
%          }
       \EndFor
       \State $A \leftarrow A \cup \{ C_{min} \}$
   \EndIf
   \EndIf
%   }
   \EndWhile

 \Procedure{$t$-\textbf{Centrality-Node-Contraction}}{$G, t$}
  \For {$(i \leftarrow 1$ to $t)$ }
  \State Select node $u$ with minimum centrality
   \If {($u$  is not cut vertex)}
     \State $V \leftarrow V \setminus \{u\}$
     \State $E \leftarrow E \setminus \{(u,v) | (u,v) \in E, \forall v \in G \}$
    \EndIf
  \EndFor
  \State \textbf{return} $G$
 %\Return $G$
 \EndProcedure

\end{algorithmic}
\end{algorithm}

\textbf{Proposition 1.}
 $t$-Centrality-Graph-Edit-Distance algorithm performs inexact GM of $G_1'$ and $G_2'$. \\

Using the properties of the edit costs of $t$-GED, the Algorithm 1 return minimum cost of complete edit path which transform input graph $G_1'$ to output graph $G_2'$, so that every vertex of $G_1'$ is uniquely corresponds to a vertex of $G_2'$. Also the procedure $t$-Centrality-Node-Contraction ensures that $V_1' \subset V_1$ and $V_2' \subset V_2$. \\

\textbf{Proposition 2.}
 The procedure $t$-Centrality-Node-Contraction executes in $O(n)$ time. \\

We can check whether a node $u$ is a cut vertex in $O(n)$ time. Therefore the \textit{for} loop of the procedure takes $O(t.n)$ time, that is $O(n)$. \\

The worst case computational complexity of the $t$-Centrality-Graph-Edit-Distance algorithm is exponential in the number of vertices in input graphs. We can use an appropriate variable $t$ to minimize the overall computation time.

\section{Experimental evaluation}
In this section, we apply $t$-Centrality-Graph-Edit-Distance algorithm for inexact GM using the degree, betweenness, eigenvector and PageRank centrality. We use IAM graph database \cite{Riesen and Bunke 2008} for the comparison of execution time and accuracy obtained by these centrality techniques. We use letter and AIDS dataset for the evaluation of the proposed inexact GM scheme.

Letter dataset contains fifteen capital letters of English alphabets, written through straight lines. For each instance of a graph, deformation of three distinct and increasing levels are applied to construct low, medium and high samples of graph dataset. Every vertex of letter graphs have an associated $(x, y)$ coordinates in the two-dimensional plane. Letter graphs with high distortion level contain the average number of vertices as 4.7 and the average number of links is 4.5.  AIDS dataset consists of graph specifying chemical compounds. It contains two class of molecules, confirmed active and confirmed inactive. Graph molecules in active class exhibit activity against HIV, whereas molecules of inactive class show inactivity against HIV. Labels on node represent chemical symbol whereas labels on edge denote valence. The average number of vertices per graph in AIDS dataset is 15.7, whereas the average number of links are 16.2 edges.

\subsection{Execution time comparison}
For the comparison purpose, we have used the value of $t^*$ in $t^*$-GED to be equal to the number of nodes which would be considered for contraction in $k^*$-degree node contraction. Therefore the value of $1^*$ in $1^*$-GED is the number of nodes of degree 1, value of $2^*$ in $2^*$-GED is the number of nodes of degree 1 followed by degree 2, similarly the value of $3^*$ in $3^*$-GED is the number of nodes of degree 1 followed by degree 2 and degree 3. Comparison of the average execution time of GM in milliseconds using $t$-Centrality-Graph-Edit-Distance algorithm as applied to letter A and E of high distortion letter dataset using different centrality measures in shown in Fig.1 and Fig.2 respectively.

\begin{figure}
\centering
%  \vspace{2.5cm}
% \includegraphics[width=0.70\textwidth]{figure/letter-a-time.eps}
 \includegraphics[width=0.70\textwidth]{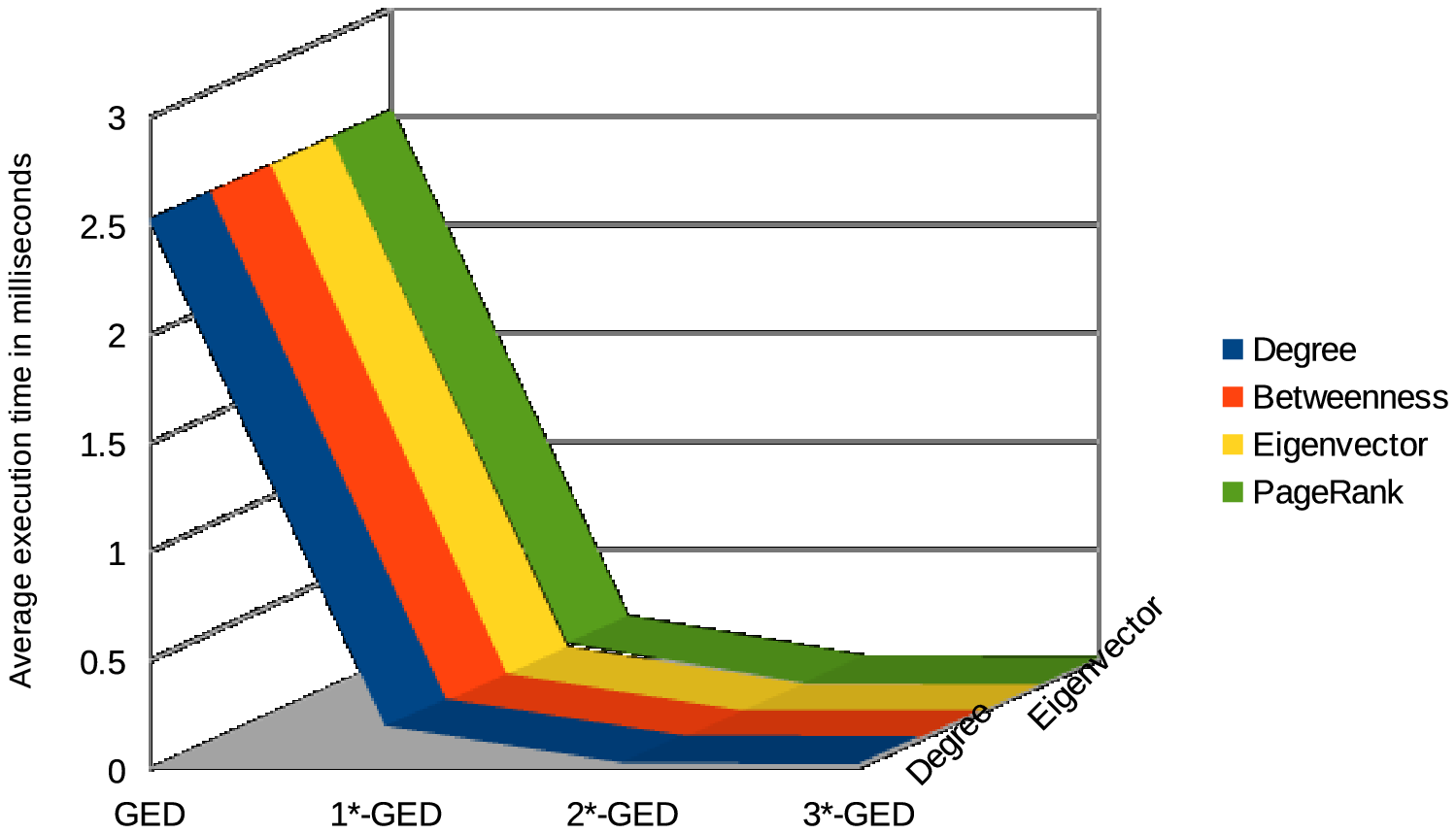}
 \caption{Comparison of execution time for letter A dataset}
\end{figure}

We can observe that GM time using eigenvector criteria is least, whereas time using degree centrality is higher. Computation time for letter E is higher as it contains more nodes than letter A.
\begin{figure}
\centering
%  \vspace{2.5cm}
% \includegraphics[width=0.70\textwidth]{figure/letter-e-time.eps}
 \includegraphics[width=0.70\textwidth]{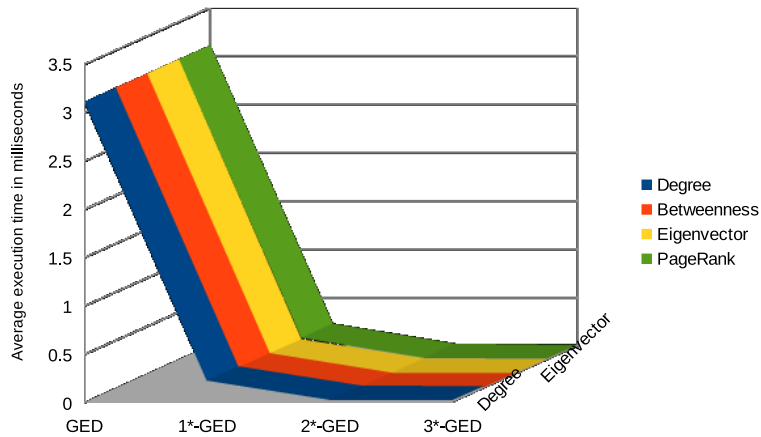}
 \caption{Comparison of execution time for letter E dataset}
\end{figure}
Comparison of the average running time of GM in milliseconds using beam search heuristic (beam width $w=10$) for the four different centrality measures for the active class of AIDS dataset are shown in Fig.3. From this figure, we observe that Algorithm 1 usually takes less time using eigenvector and betweenness centrality as compared to the degree and PageRank centrality.

\begin{figure}
\centering
%  \vspace{2.5cm}
% \includegraphics[width=0.70\textwidth]{figure/aids-active-time.eps}
 \includegraphics[width=0.70\textwidth]{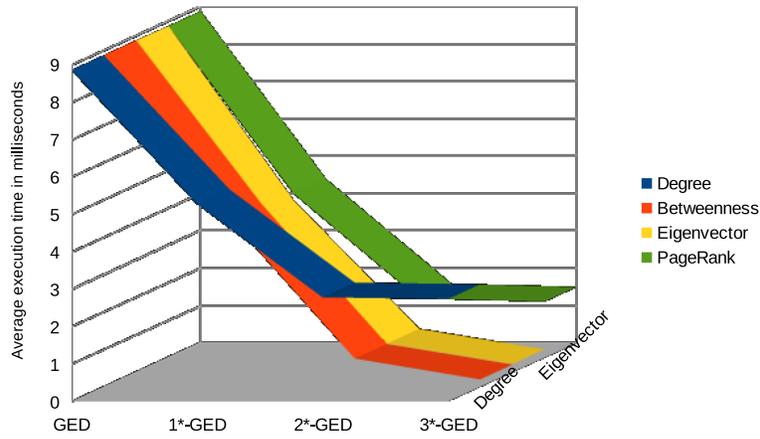}
 \caption{Comparison of execution time for active class of AIDS dataset}
\end{figure}

Fig.4 shows the corresponding average execution time of graphs for inactive AIDS dataset using the four centrality measures. Here again, the computation time using eigenvector and betweenness criteria take less time than the degree and PageRank, and between these two the average time using PageRank is less than degree centrality.

\begin{figure}
\centering
%  \vspace{2.5cm}
% \includegraphics[width=0.70\textwidth]{figure/aids-inactive-time.eps}
 \includegraphics[width=0.70\textwidth]{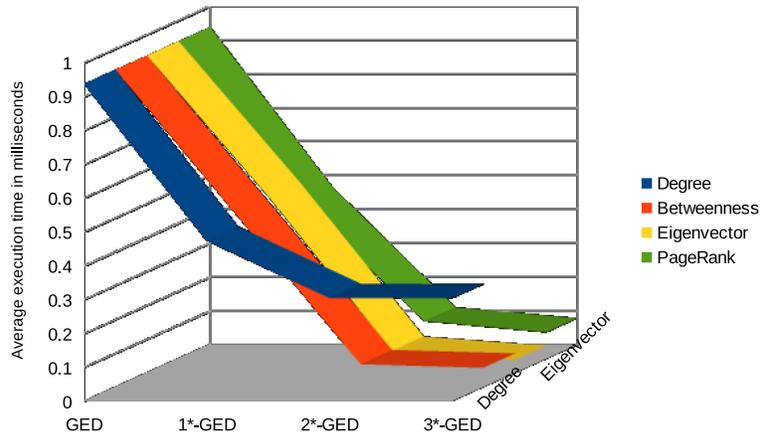}
 \caption{Comparison of execution time for inactive class of AIDS dataset}
\end{figure}

\subsection{Accuracy comparison}
For accuracy assessment, we consider the problem of classification of graphs by the nearest neighbor classifier. Letter dataset of high distortion level consists of 750 graphs for both training as well as test sets.  Each of these training, as well as test dataset, contains 50 graphs for every 15 letters. Classification accuracy of proposed GM for letter A of high distortion using the four centrality indicators is given in Fig.5, while the accuracy of GM for letter E for the same measures are shown in Fig.6.

\begin{figure}
\centering
%  \vspace{2.5cm}
% \includegraphics[width=0.70\textwidth]{figure/letter-a-accuracy.eps}
 \includegraphics[width=0.70\textwidth]{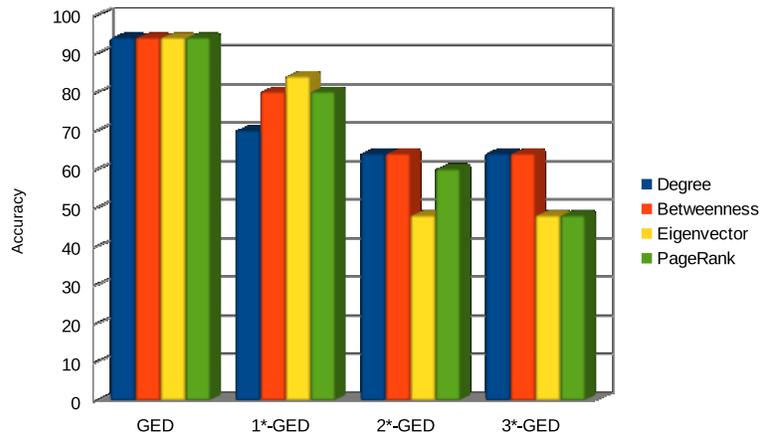}
 \caption{Comparison of accuracy ratio of letter A dataset}
\end{figure}

Here we note that accuracy of letter A for degree centrality is lower than the other three measures by contracting $t$ nodes, where t is equal to nodes with degree 1 in the input graphs ($1^*$-GED). We can also observe that for letter E, the accuracy ratio using betweenness and PageRank is usually higher than that of degree centrality even though they take less computation time.

\begin{figure}
\centering
%  \vspace{2.5cm}
% \includegraphics[width=0.70\textwidth]{figure/letter-e-accuracy.eps}
 \includegraphics[width=0.70\textwidth]{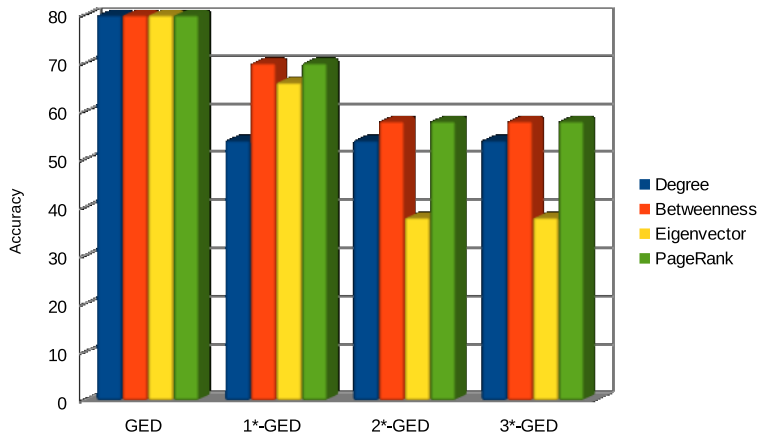}
 \caption{Comparison of accuracy ratio of letter E dataset}
\end{figure}

To find the accuracy on AIDS dataset, we utilize test dataset consisting of 300 graphs from active class and 1200 graphs from inactive class, whereas training dataset consists of 50 graphs from active class and 200 graphs from the inactive class of AIDS dataset. We can observe the accuracy ratio of the proposed inexact scheme using the four different centrality measure in Fig.7. In this figure, we observe that the accuracy obtained using degree and PageRank centrality are generally higher than that of eigenvector and betweenness centrality. Here we notice the time versus accuracy trade-off, the centrality criteria which takes less time leads to less accuracy, whereas the centrality techniques which are more accurate take more computation time.
\begin{figure}
\centering
%  \vspace{2.5cm}
% \includegraphics[width=0.70\textwidth]{figure/aids-active-accuracy.eps}
 \includegraphics[width=0.70\textwidth]{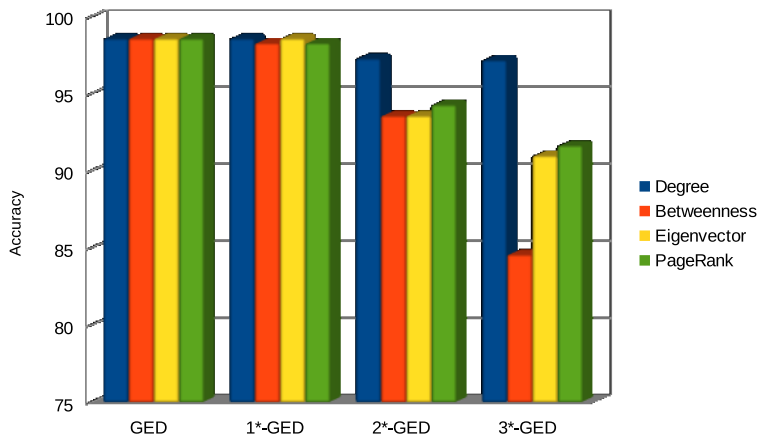}
 \caption{Comparison of accuracy for active class of AIDS dataset}
\end{figure}

Fig.8 shows the comparison of accuracy for the inactive class of AIDS dataset using the four centrality measures. In this figure also degree and PageRank criteria lead to higher accuracy for the classification of graphs of AIDS dataset.
\begin{figure}
\centering
%  \vspace{2.5cm}
% \includegraphics[width=0.70\textwidth]{figure/aids-inactive-accuracy.eps}
 \includegraphics[width=0.70\textwidth]{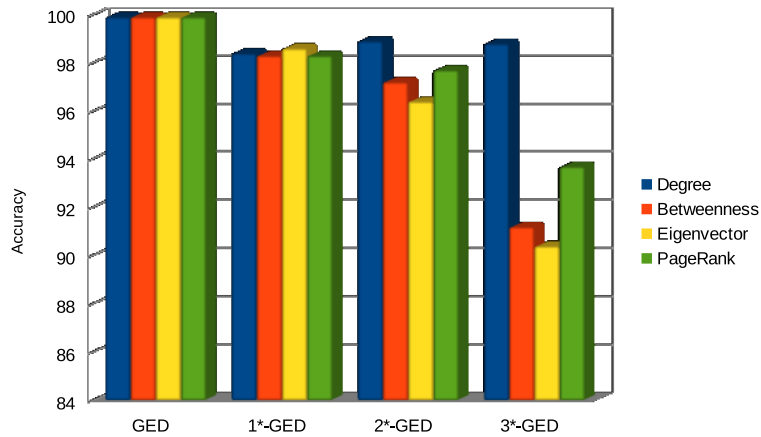}
 \caption{Comparison of accuracy for inactive class of AIDS dataset}
\end{figure}

\section{Conclusion}
In this chapter, we presented a technique to approximate GM utilizing the concept of centrality measure to reduce the size of the graphs by ignoring the nodes with a lower value of given centrality criteria. In particular, we have used eigenvector, betweenness and PageRank centrality apart from degree centrality to perform the node contraction for the computation for inexact GM. Experimental results show that these centrality criteria can be used as computation time versus accuracy trade-off for different graph dataset.

% \section*{Acknowledgment}
% more thanks here

\end{document}